\documentclass[aps,prd,showpacs,preprintnumbers]{revtex4}
\usepackage{graphics} 
\def \beq{\begin{equation}}
\def \eeq{\end{equation}}
\def \beqa{\begin{eqnarray}}
\def \eeqa{\end{eqnarray}}
\def \cs{C_s}
\def \cv{C_{\scriptscriptstyle V}}
\def \C{{\cal C}}
\def \Ds{D_s}
\def \Dt{D_t}
\def \pt{p_{\scriptscriptstyle T}}
\def \real{{\rm Re}\,}
\def \tr{{\rm Tr}\,}
\def \Z{{\cal Z}}
\def \lambdams{\Lambda_{\overline{\scriptscriptstyle MS}}}
\def \ie{{\sl i.e.\/}}
\def \etal{{\sl et al.\/}}
\def \jhep{{\sl J.\ H.\ E.\ P.\/}}
\def \np{{\sl Nucl.\ Phys.\/}}
\def \pl{{\sl Phys.\ Lett.\/}}
\def \pr{{\sl Phys.\ Rev.\/}}
\def \prl{{\sl Phys.\ Rev.\ Lett.\/}}

\begin{document}
 
\title{The speed of sound and specific heat in the QCD plasma:\\
       hydrodynamics, fluctuations and conformal symmetry}
\author{Rajiv \ V.\ \surname{Gavai}}
\email{gavai@tifr.res.in}
\affiliation{Department of Theoretical Physics, Tata Institute of Fundamental
         Research,\\ Homi Bhabha Road, Mumbai 400005, India}
\affiliation{Fakult\"at f\"ur Physik, Universit\"at Bielefeld, 
         D-33615, Germany.}
\author{Sourendu \surname{Gupta}}
\email{sgupta@tifr.res.in}
\affiliation{Department of Theoretical Physics, Tata Institute of Fundamental
         Research,\\ Homi Bhabha Road, Mumbai 400005, India.}
\author{Swagato \surname{Mukherjee}}
\email{swagato@tifr.res.in}
\affiliation{Department of Theoretical Physics, Tata Institute of Fundamental
         Research,\\ Homi Bhabha Road, Mumbai 400005, India.}

\begin{abstract}
We report the continuum limits of the speed of sound, $\cs$, and the
specific heat at constant volume, $\cv$, in quenched QCD at temperatures
of $2T_c$ and $3T_c$. At these temperatures, $\cs$ is within 2-$\sigma$
of the ideal gas value, whereas $\cv$ differs significantly from the
ideal gas. However, both are compatible with results expected in a
theory with conformal symmetry. We investigate effective measures of
conformal symmetry in the high temperature phase of QCD.
\end{abstract}
\pacs{12.38.Aw, 11.15.Ha, 05.70.Fh}
\preprint{TIFR/TH/04-33,hep-lat/0412036}
\maketitle

\section{Introduction}\label{sc.intro}

It is now a well-established fact that the pressure, $P$, and the energy
density, $\epsilon$, deviate \cite{boyd} in the high temperature
phase of QCD by about 25\% from their ideal gas values at a temperature of
about $3T_c$, where $T_c$ is the transition temperature.  Early
expectations that $\epsilon$ would count the number of degrees of
freedom in the QCD plasma through the Stefan-Boltzmann law are belied by
the fact that perturbation theory has had great difficulty in
reproducing these lattice results \cite{pert}. There have been many
suggestions for the physics implied by the lattice data, the inclusion
of various quasi-particles \cite{peshier,shuryak}, the necessity of
large resummations \cite{bir}, and effective models \cite{pisarski},
Interestingly, there has been a suggestion that conformal field theory
comes closer to the lattice result \cite{gubser}. This assumes more
significance in view of the fact that a bound on the ratio of the shear
viscosity and the entropy density, $s$, conjectured from the AdS/CFT
correspondence \cite{son} lies close to that inferred from analysis of
RHIC data \cite{teaney1} and its direct lattice measurement \cite{naka}
as well as the lattice results of a different transport coefficient
\cite{phot}.

In this paper we go beyond the measurement of the equation of state (EOS)
to thermodynamic fluctuation measures. In the pure gluon gas there is
only one fluctuation measure, the specific heat at constant volume,
$\cv$. Related to this is a kinetic variable, the speed of sound, $\cs$.
We report measurements of both of these in the high temperature phase of
the pure glue plasma, through a continuum extrapolation of results
obtained with successively finer lattices. Not only do these quantities
provide further tests of all the models and expansions which try to
explain the lattice data on the EOS, they also have direct physical
relevance to ongoing experiments at the RHIC in Brookhaven.

The speed of sound governs the evolution of the fire-ball produced
in the heavy-ion collision and  hence plays a crucial role in the
hydrodynamic study of the signatures of QGP.  Elliptic flow is one of
the most important quantity that has been suggested for the signature
of QGP formation in the heavy-ion collision experiments. It has
been shown \cite{ollitrault,sorge, kolb,teaney} that elliptic flow is
sensitive to the value of $\cs$.

In recent years, event-by-event fluctuation of quantities has been of
immense interest as signatures of quark-hadron phase transition. In
order to use fluctuations as a probe of the plasma phase, one has to
identify observables whose fluctuations survive the freeze-out of the
fireball. The evolution of these fluctuations is sensitive to the values
of $\cs$, as shown in \cite{mohanty} for the case of net baryon number
fluctuation. In \cite{gazdzicki} it has been claimed that, within the
regime of thermodynamics, the ratio of the event-by-event fluctuations
of entropy and energy is given by
\beq
   R_e=\frac{(\delta S)^2/S^2}{(\delta E)^2/E^2}=\frac1{(1+\cs^2)^2},
\eeq 
and hence provides an estimate of the speed of sound, if $R_e$ turns out
to be measurable in heavy-ion collisions.

The specific heat is, of course, directly a measure of fluctuations.
It was suggested in \cite{stodolsky} that event-by-event temperature
fluctuation in the heavy-ion collision experiments be used to measure
$\cv$. Also it has been shown in \cite{korus} that $\cv$ is directly
related to the event-by-event transverse momentum ($\pt$) fluctuations.

The measurement of $\cv$ and $\cs$ also directly test the relevance of
conformal symmetry to finite temperature QCD. It is common knowledge
that QCD generates a scale, $\Lambda_{QCD}$, microscopically, and
thus breaks conformal invariance. The strength of the breaking of this
symmetry at any scale is parametrized by the $\beta$-function. A recent
suggestion is that an effective theory which reproduces the results of
thermal QCD at long-distance scales may somehow be close to a conformal
theory. The result of \cite{gubser} for the entropy density, $s$,
in a Yang-Mills theory with four supersymmetry charges (${\cal N}=4$
SYM) and large number of colours, $N_c$, at strong coupling, is
\beq
   \frac s{s_0} = f(g^2 N_c), \quad{\rm where}\quad
   f(x)=\frac34+\frac{45}{32}\zeta(3)(2x^{-3/2})+\cdots
   \quad{\rm and}\quad s_0 = \frac23 \pi^2 N_c^2T^3,
\label{sym}\eeq
where $g$ is the Yang-Mills coupling.  For the $N_c=3$ case at hand,
the well-known result for the ideal gas, $s_0=4(N_c^2-1)\pi^2T^3/45$
takes into account, through the factor $N_c^2-1$, the relatively important
difference between a $SU(N_c)$ and an $U(N_c)$ theory.

Of course, at finite temperature there is a scale, $T$, which appears in,
for example, $\epsilon$ as a factor of $T^4$. However, the strength of the
breaking of conformal symmetry must be measured as always, through the
trace of the stress-tensor. After subtracting the ultraviolet divergent
($T=0$) pieces, this is given by the so-called interaction measure,
$\Delta=\epsilon-3P$.  Thus the ratio, $\C=\Delta/\epsilon$, which we
call the conformal measure, parametrizes the departure from conformal
invariance at the long-distance scale \footnote{It is equally possible to
write a temperature dependent gluon condensate from the trace of the full
stress-energy tensor. However, that trace requires renormalization. After
proper renormalization, its value is not $\Delta$, but $\Delta+
B(g)\langle F^2\rangle_0/2g^3$, where $\langle F^2\rangle_0$ is the
gluon condensate at $T=0$.}. We shall present data for this
quantity in this paper. Note that when $\Delta=0$ one has $\cs^2=1/3$
and $\cv/T^3=4\epsilon/T^4$. In addition, if the gas is ideal, then one
obtains $\epsilon/T^4=(N_c^2-1)\pi^2/15$.

The paper is organized as follows. In the next section we present the
formalism and lead up to the measurement of $\cv$ and $\cs^2$
on the lattice in Section \ref{sc.cvcs}. Some details are given in
the appendices.  In Section \ref{sc.results} we give details of our
simulations and our results. Finally, in Section \ref{sc.summary}
we present a discussion of the results. Those who are interested only
in the results may read the initial part of Section \ref{sc.formalism}
and then jump to Section \ref{sc.summary}.

\section{Formalism} \label{sc.formalism}

For any theory on the lattice one may compute derivatives of the partition function,
$\Z(V,T)$, where $V$ is the volume and $T$ the temperature. In
particular the energy density, $\epsilon$, and the pressure, $P$, are
given by the first derivatives of $\ln\Z$,
\beq
   \epsilon = \frac{T^2}{V}
     \left.\frac{\partial\ln\Z(V,T)}{\partial T}\right|_V,
   \qquad{\rm and}\qquad
   P = T\left.\frac{\partial\ln\Z(V,T)}{\partial V}\right|_T.
\eeq
The second derivatives are measures of fluctuations. For a relativistic gas,
where particles may be created and destroyed, there is only one second derivative
in the absence of chemical potentials, the specific heat at constant volume---
\beq
   \cv = \left.\frac{\partial\epsilon}{\partial T}\right|_V.
\label{specht} \eeq
A quantity such as the compressibility is not defined in the absence of a chemical
potential, since a gas may adapt to a change in volume without change in pressure
by simply creating or destroying particles. Nevertheless, the speed of sound
is a well defined quantity. One can see this by using thermodynamic identities
to recast the definition in the form
\beq
   \cs^2 \equiv \left.\frac{\partial P}{\partial\epsilon}\right|_s
     = \left.\frac{\partial P}{\partial T}\right|_V
      \left(\left.\frac{\partial\epsilon}{\partial T}\right|_V\right)^{-1}
     = \frac{s/T^3}{\cv/T^3},
\label{sound} \eeq
where we have used the thermodynamic identity
\beq
   \left.\frac{\partial P}{\partial T}\right|_V = 
      \left.\frac{\partial S}{\partial V}\right|_T 
   \quad{\rm and}\quad
      \left.\frac{\partial S}{\partial V}\right|_T =
    s = \frac{\epsilon+P}T,
\label{entropy} \eeq
in conjunction with the definition of the entropy density, $s$, above.
Although this work deals exclusively with the pure gauge theory, this
seems to be a good place to remark that in full QCD without quark chemical
potentials all these relations go through. However, in the presence of a
chemical potential, compressibility is a well defined concept, and the
speed of sound may be affected by this new physics. We shall deal with
this elsewhere.

\subsection{Energy density and pressure} \label{sc.eos}

In order to set up our notation, and for the sake of completeness, we write
down some results which have been known since the beginning of lattice
thermodynamics. For the pure gauge $SU(N_c)$ theory with the Wilson action,
the partition function is
\beq
   {\cal Z}(V,T)=\int{\cal{D}}U e^{-S[U]},
   \qquad{\rm where}\qquad
   S[U]=2N_c  \left[ K_{s}P_s+K_{\tau}P_{\tau} \right],
\eeq
periodic boundary conditions are assumed,
$P_s$ denotes the sum of spatial plaquettes over all lattice sites,
and $P_\tau$ is the corresponding sum of mixed space-time 
plaquettes\footnote{We define a plaquette value as $1 - \real \tr (UUUU)/N_c$
where the string of $U$'s is the product of link matrices taken in
order around a plaquette.}. If the lattice spacings, $a_i$ ($i=s,\tau$)
along spatial ($s$) and temporal ($\tau$) directions, are related by the
anisotropy parameter $\xi=a_s/a_{\tau}$, then the coupling may be written
as $K_s=1/\xi g_s^2$, and $K_{\tau}=\xi/g_{\tau}^2$. $\xi=1$ corresponds
to the usual case of symmetric lattice spacings, \ie, $a_s=a_\tau$.
In the weak coupling limit, $g_i^{-2}$ 's can be expanded \cite{Has}
around their symmetric lattice value $g^{-2}(a)$,
\beq 
   g_i^{-2}(a_s,\xi)=g^{-2}(a)+c_i(\xi)+O[g^2(a)],
\label{coupling2} \eeq
with the condition $c_i(\xi=1)=0$.

The computation of the first derivatives of the couplings requires
$b(a_s)=a_s\partial g^{-2}/\partial a_s$. With
the usual definition of the $\beta$-function-
\beq
   B(\alpha_s) = \frac\mu2\,\frac{\partial\alpha_s}{\partial\mu}
     = -\,\frac{33-2N_f}{12\pi}\alpha_s^2+\cdots, \qquad{\rm where}\qquad
   \alpha_s=g^2/4\pi,
\eeq
one finds $ b(a_s) = B(\alpha_s)/2 \pi \alpha_s^2$. Then, 
\beq
   a_s \frac{\partial K_s}{\partial a_s} = \frac{B(\alpha_s)}{2 \pi \alpha_s^2 \xi},\quad
   K_s' = - \frac{g_s^{-2}}{\xi^2}+\frac{c_s'}{\xi},
   \qquad{\rm and}\qquad
   a_s\frac{\partial K_{\tau}}{\partial a_s} = \frac{\xi B(\alpha_s)}{2 \pi \alpha_s^2},\quad
   K_{\tau}' = g_{\tau}^{-2} + \xi c_{\tau}',
\label{coup-der-1} \eeq
where primes denote derivative with respect to $\xi$. The quantities
$c_s'$ and $c_\tau'$ have been computed to one-loop order in the weak
coupling limit for $SU(N_c)$ gauge theories \cite{Kar}.
\par
If the number of lattice sites in the two directions are $N_\tau$ and $N_s$, then
$T=1/N_\tau a_\tau$ and $V=(N_s a_s)^3$. The partial derivatives with respect
to $T$ and $V$ can be written in terms of the two lattice parameters $\xi$ and $a_s$,
keeping $N_s$ and $N_\tau$ fixed,
\beq
   T\left.\frac{\partial}{\partial T}\right|_{V} \equiv
       \xi\left.\frac{\partial}{\partial \xi}\right|_{a_s},
   \qquad{\rm and}\qquad
   3V\left.\frac{\partial}{\partial V}\right|_{T} \equiv
       a_s\left.\frac{\partial}{\partial a_s}\right|_{\xi} +
       \xi\left.\frac{\partial}{\partial \xi}\right|_{a_s}.
\label{lat-der} \eeq
Using these derivatives, one obtains \cite{EKS1}
\beq
   a_s^4\epsilon = -6 N_c \xi^2 \left[ \frac{\partial K_s}{\partial \xi}\Ds
           +\frac{\partial K_{\tau}}{\partial\xi}\Dt \right],
   \quad{\rm and}\quad
   a_s^4\Delta \equiv a_s^4\left(\epsilon-3P\right)
      = 6 N_c \xi a_s \left[ \frac{\partial K_s}{\partial a_s}\Ds
           +\frac{\partial K_{\tau}}{\partial a_s}\Dt \right].
\label{pressure} \eeq
A subtraction of the corresponding vacuum ($T=0$) quantities lead to 
$D_i=\langle \bar{P_i} \rangle -\langle \bar{P_0} \rangle$ above,
where $\langle \bar{P_0}\rangle$ is the average plaquette value at $T=0$,
and the bar denotes division by $3N_s^3N_\tau$. The $T=0$ quantities are
evaluated by simulations on symmetric lattices with $N_\tau=N_s\to\infty$. 
In the isotropic limit, $\xi=1$, one obtains
\beq
   \frac\epsilon{T^4}=6N_cN_\tau^4 \left[\frac{\Ds-\Dt}{g^2}
         -(c_s'\Ds+c_{\tau}'\Dt) \right],
   \qquad{\rm and}\qquad
   \frac\Delta{T^4}=6N_cN_\tau^4 \frac{B(\alpha_s)}{2 \pi \alpha_s^2} \biggl[\Ds+\Dt\biggr].
\label{pressure1} \eeq
Note that $\Delta$ contains $B(\alpha_s)$ as a factor, but this explicit 
breaking of conformal symmetry may be compensated by the vanishing of the 
factor $\Ds+\Dt$.  To determine the coupling $g^2$, throughout this work,
we use the method suggested in \cite{sgupta}, where the one-loop order
renormalized couplings have been evaluated by using $V$-scheme \cite{LM}
and taking care of the scaling violations due to finite lattice spacing
errors using the method in \cite{EHK}.

\subsection{The specific heat and speed of sound} \label{sc.cvcs}

\begin{figure}
\begin{center}
   \scalebox{1.0}{\includegraphics{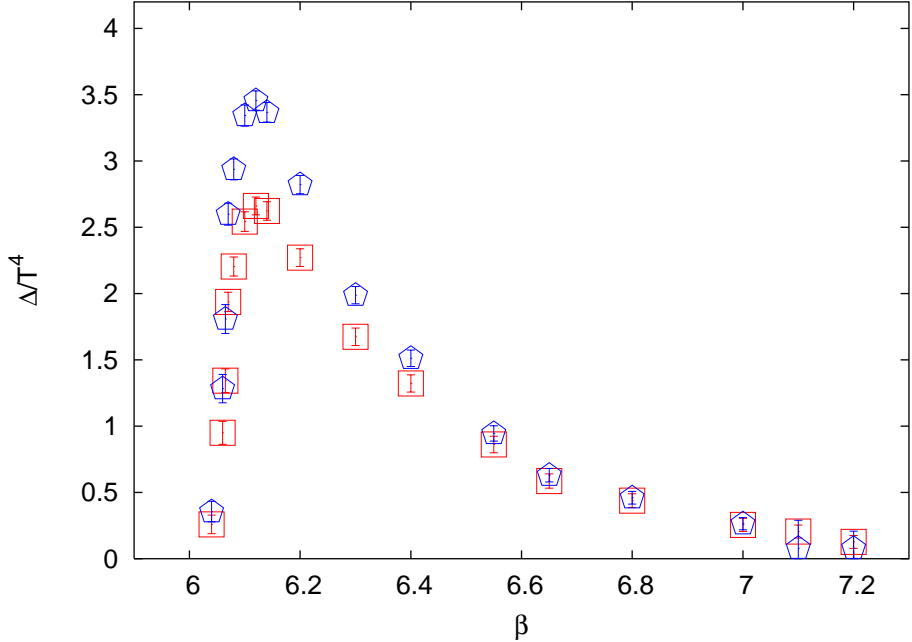}}
\end{center}
\caption{$\Delta/T^4$ as a function of the bare coupling $\beta$ using a
   non-perturbative (squares) and one-loop order perturbative (pentagons)
   $\beta$-function, $B(\alpha_s)$. The results agree for $\beta\ge6.5$. The data
   of \cite{boyd} for $N_\tau=8$ has been used in this analysis.}
\label{fg.D-int-diff} \end{figure}

Direct application of the derivatives in eq.\ (\ref{lat-der}) to the
expression for $\epsilon$ in eq.\ (\ref{pressure}) would give an incorrect
result for $\cv$, \ie , $\cv/T^3$ would not give the correct ideal gas
value ($4\epsilon/T^4$) in the limit $g \rightarrow 0$. This is because
we have chosen to work with the variables $\xi$ and $a_s$, whereby the
dimensions of both $T$ and $V$ come from powers of $a_s$. Application
of the derivative formul\ae{} therefore see false scalings of these
quantities.  We solve this problem by choosing to work in terms of a
dimensionless ratio so that the scaling is automatically taken care
of. We define
\beq
   \C = \frac\Delta\epsilon \qquad{\rm and}\qquad
   \Gamma = T\left.\frac{\partial\C}{\partial T}\right|_V,
\label{defgamma} \eeq
where $\C$ is the conformal measure already discussed. Then, using eqs.\ 
(\ref{entropy}, \ref{defgamma}) one can proceed straightaway to write
\beq
   \frac\cv{T^3} = \left(\frac{\epsilon/T^4}{P/T^4}\right)
     \left[\frac s{T^3}+\frac{\Gamma}{3} \frac\epsilon{T^4}\right],
   \qquad{\rm and}\qquad
   \cs^2= \left(\frac{P/T^4}{\epsilon/T^4}\right)
      \left[ 1+\frac{\Gamma\epsilon/T^4}{3s/T^3}\right]^{-1}.
\label{CvCs} \eeq
In order to complete these expression, one needs to express $\Gamma$
in terms of quantities computable on the lattice.

We lighten the subsequent formul\ae{} by defining the two functions
\beq
   F(\xi, a_s) = \frac{\Delta a_s^4}{6N_c\xi} =
     a_s \left[ \frac{\partial K_s}{\partial a_s}\Ds
       +\frac{\partial K_{\tau}}{\partial a_s}\Dt \right],
   \qquad{\rm and}\qquad
   G(\xi, a_s) = \frac{-\epsilon a_s^4}{6N_c\xi} =
     \xi \left[ \frac{\partial K_s}{\partial\xi}\Ds
       + \frac{\partial K_{\tau}}{\partial\xi}\Dt \right].
\label{F-G} \eeq
Since $\C=-F/G$, one finds that
\beq
   \Gamma = -\frac TG \left.\frac{\partial F}{\partial T}\right|_V
        + \frac{TF}{G^2}\left.\frac{\partial G}{\partial T}\right|_V.
\label{gamma}\eeq
The derivatives of $F$ and $G$ involve the variances and covariances
of the spatial and temporal plaquettes. The complete expressions are
given in Appendix \ref{sc.variance-terms}.  We also need to write down
the second derivatives of the couplings---
\beq
   a_s\frac{\partial K_s'}{\partial a_s}=-\frac{B(\alpha_s)}{2 \pi \alpha_s^2 \xi^2},\quad
   K_s'' = \frac{2g_s^{-2}}{\xi^3}-\frac{2c_s'}{\xi^2}+\frac{c_s''}{\xi},\qquad
   a_s\frac{\partial K_\tau'}{\partial a_s}=\frac{B(\alpha_s)}{2 \pi \alpha_s^2},\quad
   K_{\tau}'' = 2c_{\tau}'+\xi c_{\tau}''.
\label{coup-der-2} \eeq
Details of the evaluation of the second order Karsch coefficients
$c_{s,\tau}''$ are given in Appendix \ref{sc.karsch-coeff}. For $N_c=3$
their numerical values are $c_s''=-0.298193$ and $c_\tau''=0.333670$.

It is known \cite{Heller} that in the weak-coupling limit, \ie, for
$g\to0$, the dominant contribution to all plaquettes varies as $g^2$. So
the $D_i$'s also vary as $g^2$.  Hence in this limit $\Delta/T^4\propto
g^2$ and can be neglected in comparison with $\epsilon/T^4$ which is
$(N_c^2-1)\pi^2/15 +{\cal O}(g^2)$.  As shown in Appendix
\ref{sc.variance-terms}, $\Gamma \to 0$ in this limit.  As a result,
$\cv/T^3\to4\epsilon/T^4$ and $\cs^2\to1/3$.  Note that in any conformal
invariant theory in $d+1$ dimensions one has $\epsilon=dP$, \ie,
$\C=\Gamma=0$, and hence, by eq.\ (\ref{CvCs}), $\cs^2=1/d$ and
$\cv/T^3=(d+1)\epsilon/T^4$.

\subsection{On the method}

The formalism outlined in Section \ref{sc.eos} corresponds to the
so-called ``differential'' method. On the other hand, about a decade back
a new method called the ``integral'' method was employed \cite{boyd}
to find the equation of state of QCD matter. This consists of using the
expression for $\Delta/T^4$ in eq.\ (\ref{pressure1}) while eschewing
the expression for $\epsilon/T^4$ in favour of a totally different
method for determining $P/T^4$. This proceeds from the observation that
$P=(T/V)\ln\Z$ for a homogeneous system. Since lattice computations do
not determine the absolute normalization of $\Z$, use of this formula
requires fixing an additive constant. In \cite{boyd} 
this constant was fixed by setting $P=0$ in the low temperature
phase of QCD, slightly below $T_c$.

This leads to a conceptual problem. Since the pure gauge phase transition
in QCD is of first order, the system is not homogeneous at $T_c$
and the method is not strictly applicable there. One makes an unknown
systematic error in integrating through $T_c$. This is in addition to a
small systematic error due to setting $P=0$ just below $T_c$
and the numerical integration errors. However, a
decade ago, these errors were deemed to be smaller than the scale errors
in the ``differential'' method which could give rise even to negative
pressure in computations with coarse lattice spacings, $a=1/TN_\tau$, which
were the only ones possible then.

The two methods must agree in the continuum limit. A reanalysis of the
data of \cite{boyd} showed that the two methods agreed to within 5\%
(\ie, within statistical errors) for $T\ge2T_c$ \cite{sgupta} already for
$N_\tau=8$, \ie, for $a=1/16T_c$. In fact a criterion for the agreement is
straightforward, and follows from the fact the expression for $\Delta/T^4$
is common to the two methods. Since the integral method does not need
the Karsch coefficients, it allows one to use a non-perturbatively
determined $\beta$-function $B(\alpha_s)$. On the other hand, the differential
method requires, for internal consistency, that the Karsch coefficients
and $B(\alpha_s)$ be obtained at the same order, \ie, at one-loop order in
the present state of the art. 

Thus, a comparison between the values of $\Delta/T^4$ extracted for a
given $N_\tau$ using the two techniques would reveal at what $T$ the two
methods become identical. Then, using asymptotic scaling, one could also
give the minimum value of $N_\tau$ which would be required for the same level
of agreement as a function of $T$.  Such a comparison is shown in Figure
\ref{fg.D-int-diff} and demonstrates that a bare coupling of $\beta\ge6.55$
already suffices.

The use of the ``differential'' method does require small lattice
spacings, but this is currently quite feasible with modern computers. The
advantage is that the formalism of Section \ref{sc.cvcs} can be used to
compute second derivatives of the free energy as operator expectation
values. If the EOS were to be evaluated by the ``integral'' method, then
these quantities could only be evaluated by numerical differentiation
\cite{iitk}, which is a less attractive solution.

\begin{table}
\begin{center}
\begin{tabular}{|c|c|c c|c c|}
\hline 
$T/T_c$&$\beta$&\multicolumn{2}{c|}{Asymmetric Lattice}&\multicolumn{2}{c|}{Symmetric Lattice} \\\cline{3-6}
&&size&stat.&size&stat. \\ \hline
2.0&6.0625&$4 \times 8^3$&60000&$22^4$&38000\\
&&$~\times 10^3$&56000&&\\
&&$~\times 12^3$&50000&&\\
&&$~\times 14^3$&51000&&\\
&&$~\times 16^3$&51000&&\\
&&&&&\\
&6.5500&$8 \times 18^3$&1100000&$12^4$&800000\\
&&&&$16^4$&626000\\
&&&&$18^4$&802000\\
&&&&$22^4$&552000\\
&&&&$32^4$&330000\\
&&&&&\\
&6.7500&$10 \times 22^3$&1100000&$22^4$&969000\\
&6.9000&$12 \times 26^3$&1040000&$26^4$&550000\\
&7.0000&$14 \times 30^3$&425000&$30^4$&146000\\ \hline
3.0&6.3384&$4 \times 10^3$&220000&$22^4$&75000\\
&&$~ \times 12^3$&200000&&\\
&&$~ \times 16^3$&200500&&\\
&&$~ \times 20^3$&210000&&\\
&&$~ \times 22^3$&150000&&\\
&7.0500&$10 \times 32^3$&560000&$32^4$&146000\\
&7.2000&$12 \times 38^3$&315000&$38^4$&58000 \\\hline
\end{tabular}
\end{center}
\caption{The coupling ($\beta$), lattice sizes ($N_\tau\times N_s^3$), statistics
   and symmetric lattice sizes ($N_s^4$) are given for each temperature. Statistics
   means number of sweeps used for measurement of plaquettes after discarding for
   thermalization.} 
\label{tb.simulation} \end{table}

\begin{figure}
\begin{center}
   \scalebox{0.6}{\includegraphics{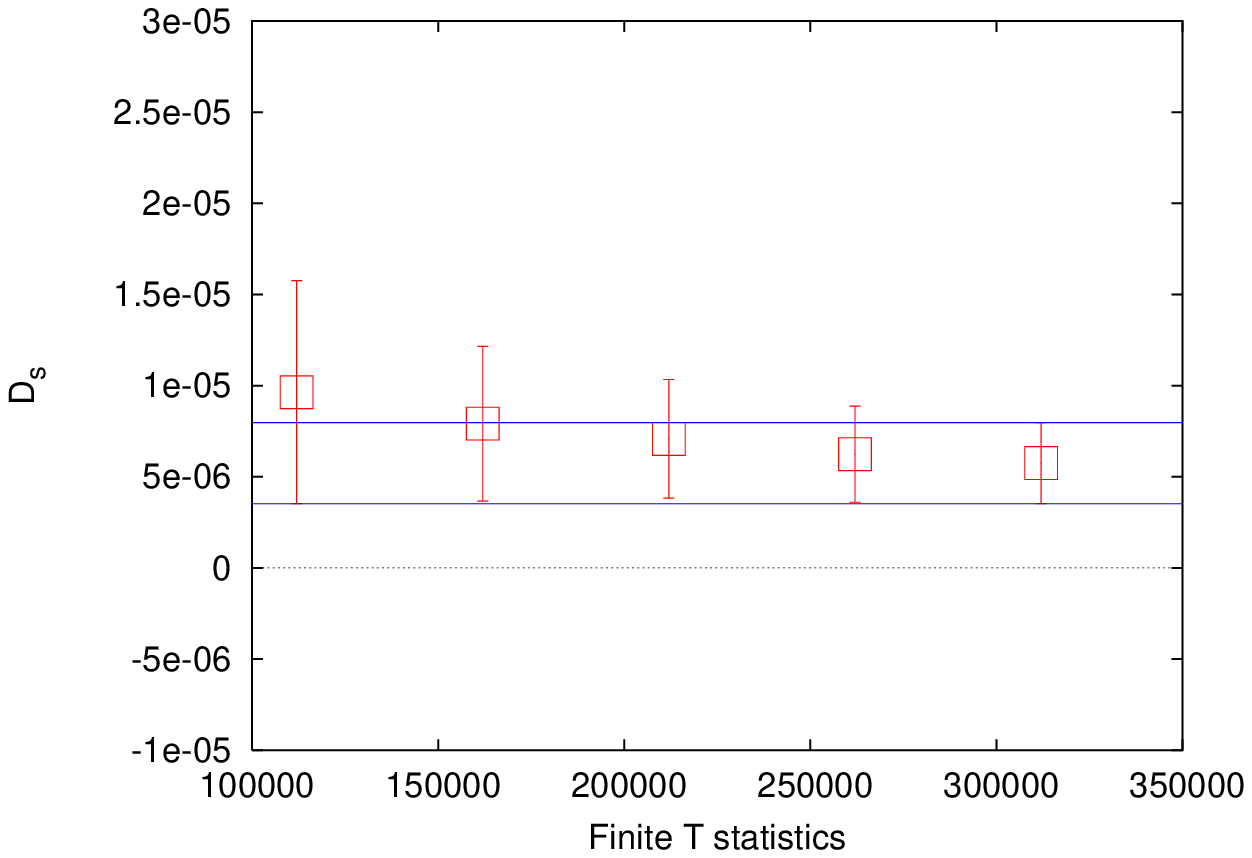}}
   \scalebox{0.6}{\includegraphics{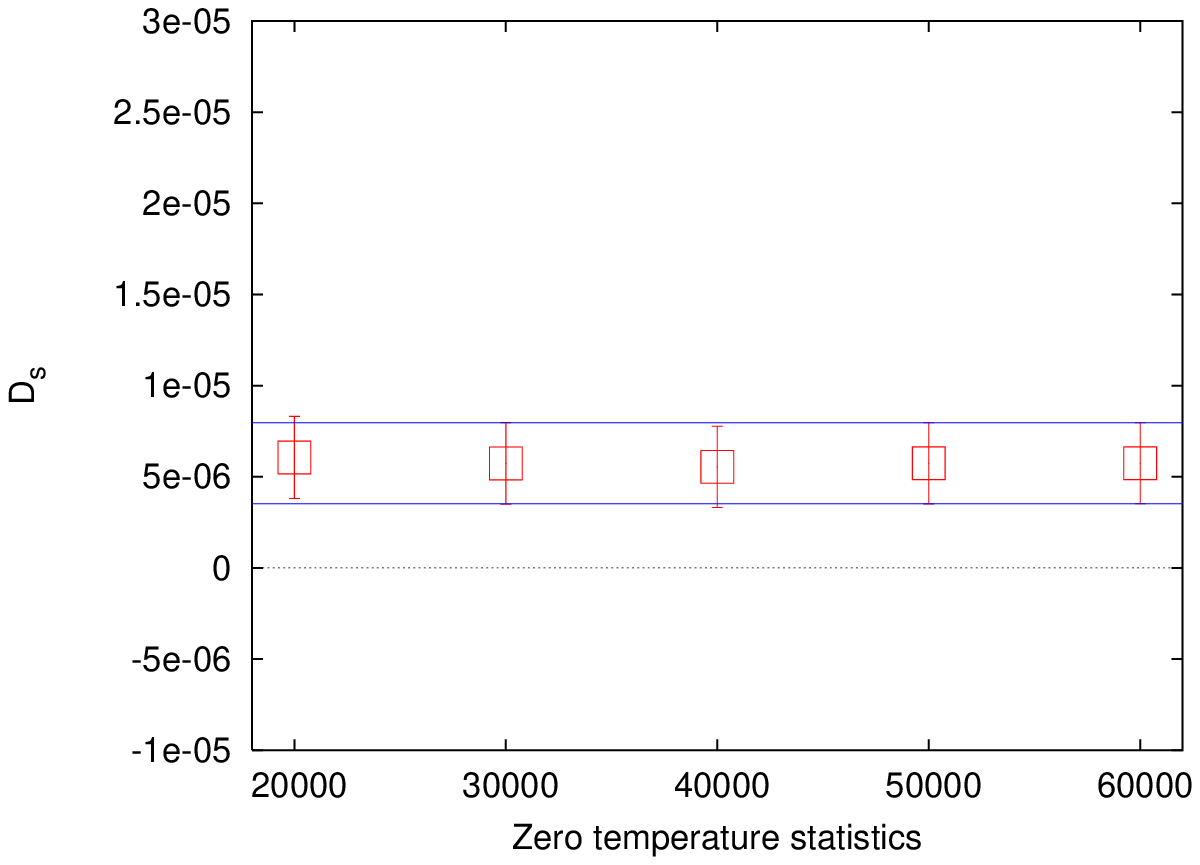}}
\end{center}
\caption{Stability of $\Ds$ against the statistics in $T>0$ (first panel)
   and $T=0$ (second panel) simulations. The coupling corresponds to
   $3T_c$ on a $12\times38^3$ lattice and the corresponding $T=0$
   simulation was performed on a $38^4$ lattice. In both the
   figures we have also plotted the 1-$\sigma$ error band of the final
   errors.}
\label{fg.Ds-stat} \end{figure}

\begin{figure}
\begin{center}
   \scalebox{0.6}{\includegraphics{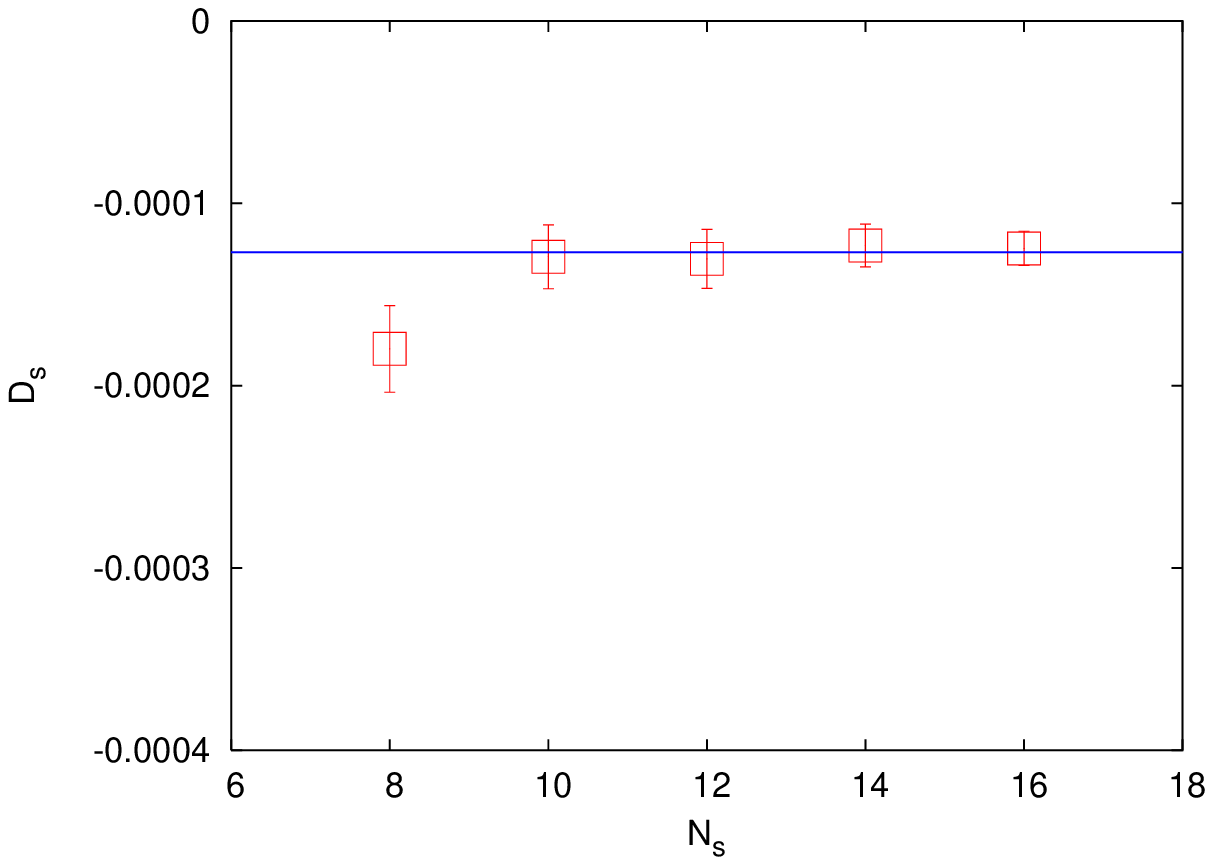}}
   \scalebox{0.6}{\includegraphics{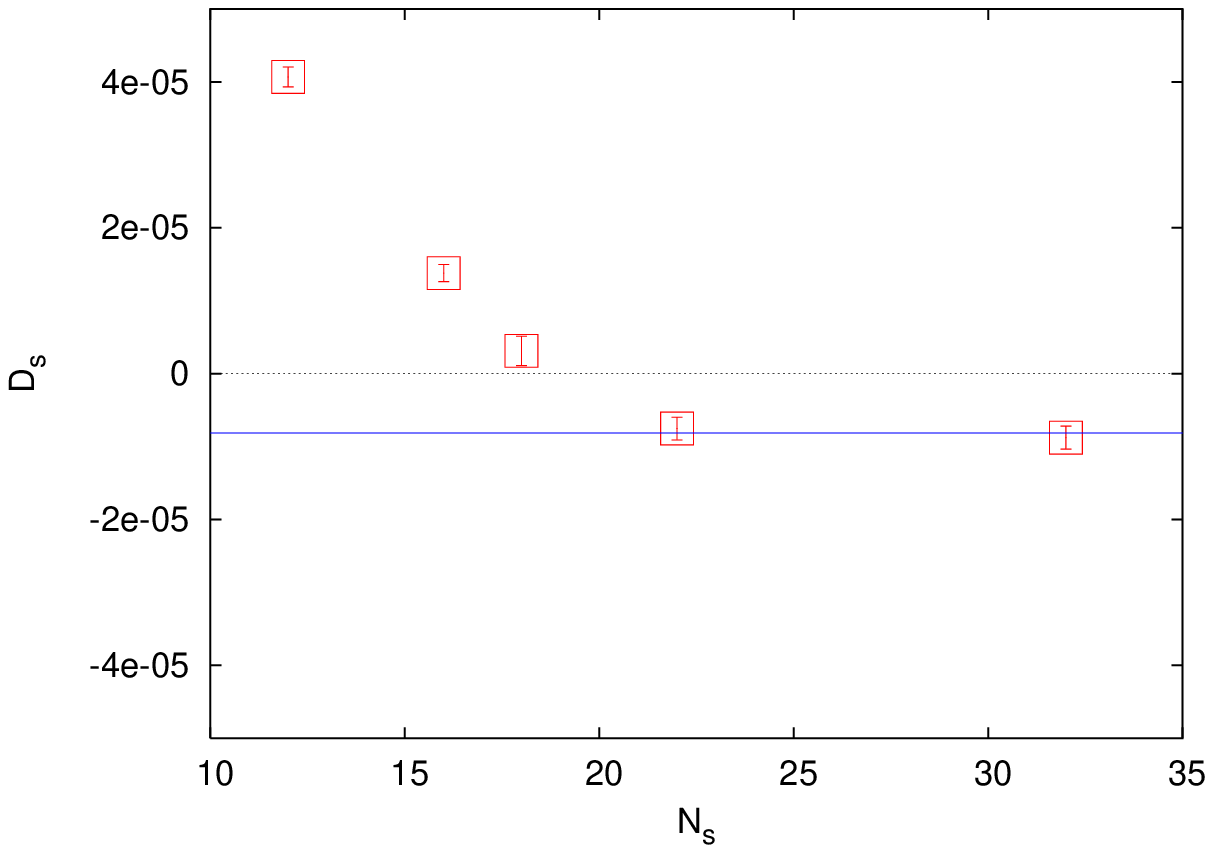}}
\end{center}
\caption{Dependence of $\Ds$ against the spatial size, $N_s$, of the $T>0$
   lattice for $2T_c$ and $22^4$ lattice for the $T=0$ computation (left
   panel). We have shown a fit to a constant through the three largest
   lattices. Dependence of $\Ds$ on $N_s$ for the $T=0$ lattice when
   the $2T_c$ computation is performed on an $8\times18^3$ lattice (right
   panel). A fit to a constant with the two largest sizes is also shown.}
\label{fg.Ds-vol} \end{figure}

\begin{table}
\begin{center}
\begin{tabular}{|c|ccc|} \hline
 $N_\tau$&$36N_c^2N_{\tau}^4V\sigma_{s,s}$&
 $36N_c^2N_{\tau}^4V\sigma_{\tau,\tau}$&$36N_c^2N_{\tau}^4V\sigma_{s,\tau}$\\
\hline 
10&$(0.2\pm1.0) \times 10^{-5}$&
   $(0.2\pm1.0) \times 10^{-5}$&
   $(0.1\pm1.0) \times 10^{-5}$ \\
12&$(0.7\pm4.1) \times 10^{-4}$&
   $(0.7\pm4.1) \times 10^{-4}$&
   $(0.4\pm3.5) \times 10^{-4}$ \\
14&$(0.1\pm1.4) \times 10^{-3}$&
   $(0.1\pm1.5) \times 10^{-3}$&
   $(0.1\pm1.3) \times 10^{-3}$ \\
\hline 
\end{tabular}
\end{center}
\caption{Contributions of different covariances, with their respective 
   errors, are tabulated for different $N_\tau$ at $2T_c$. This shows 
   all the contributions are totally negligible.}
\label{tb.covar}
\end{table}

\begin{figure}
\begin{center}
   \scalebox{0.6}{\includegraphics{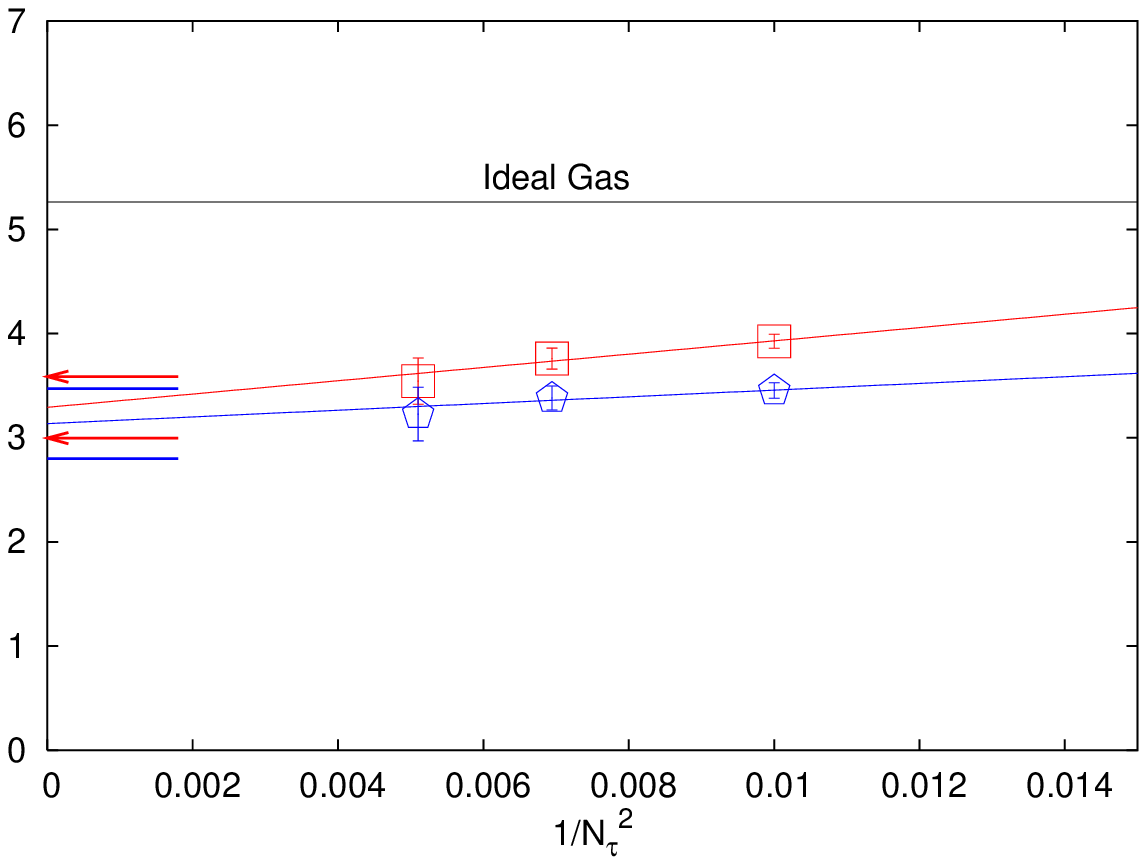}}
   \scalebox{0.6}{\includegraphics{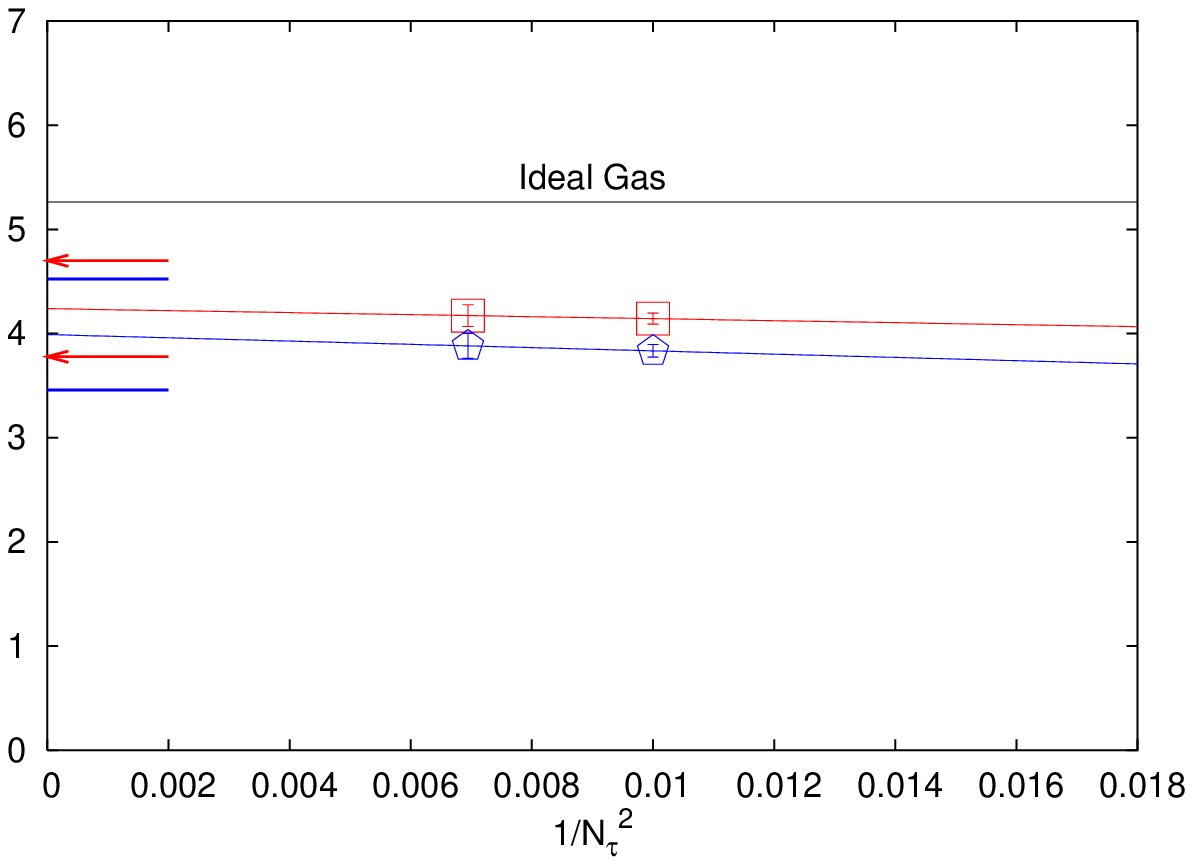}}
\end{center}
\caption{Dependence of $\epsilon/T^4$ (squares) and $3P/T^4$ (pentagons) on
   $1/N_\tau^2$ for $T=2T_c$ (left panel) and $T=3T_c$ (right panel). The
   1-$\sigma$ error band of the continuum values has been indicated by arrows
   (for $\epsilon/T^4$) and lines (for $3P/T^4$).}
\label{fg.E-P} \end{figure}

\begin{figure}
\begin{center}
   \scalebox{0.6}{\includegraphics{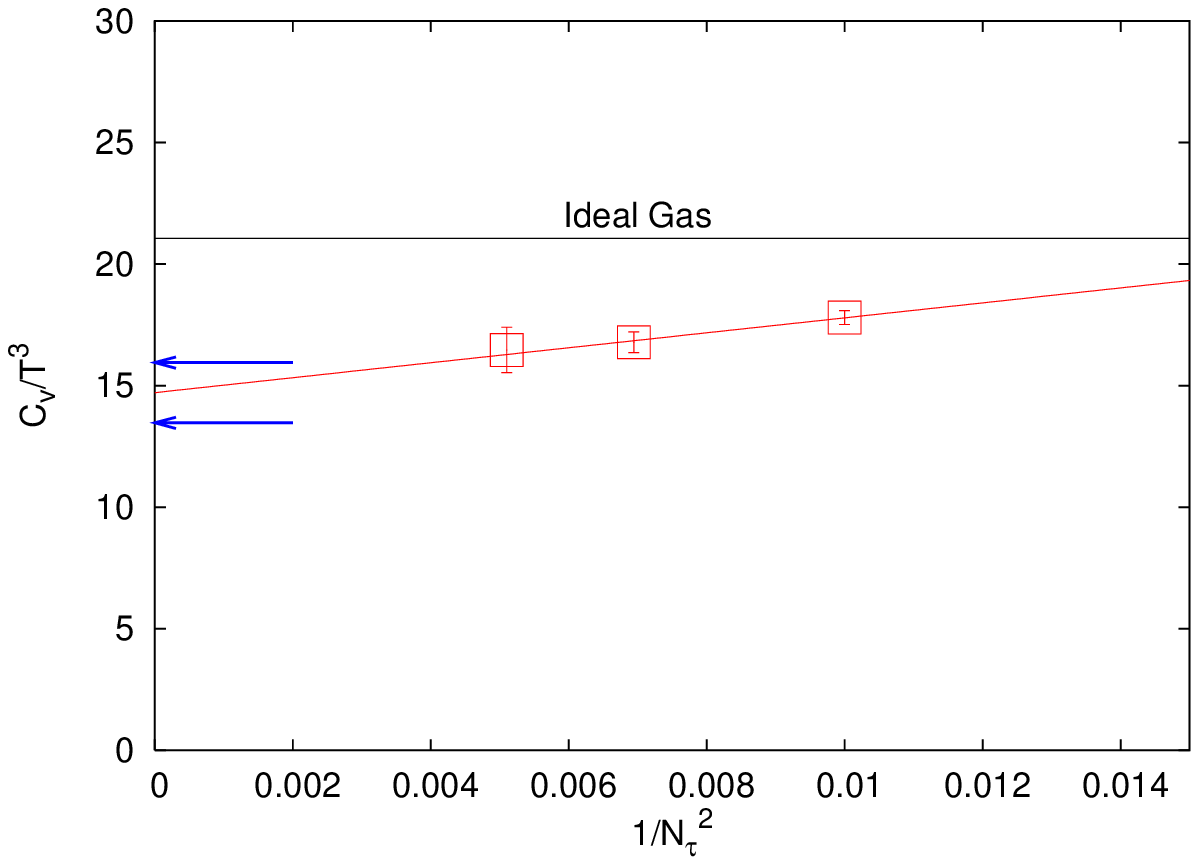}}
   \scalebox{0.6}{\includegraphics{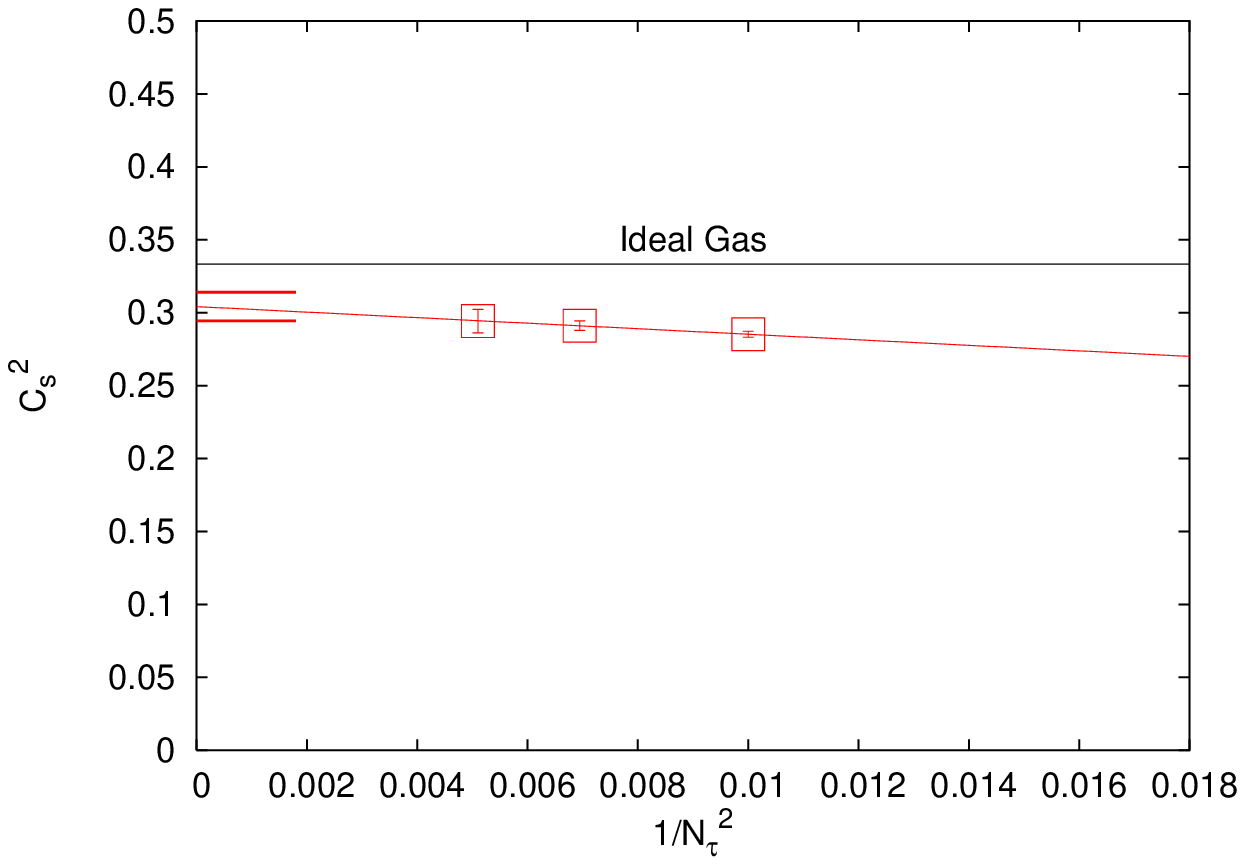}}
\end{center}
\caption{Dependence of $\cv/T^3$ (left panel) and $\cs^2$ (right panel) on
   $1/N_\tau^2$ at $T=2T_c$. The 1-$\sigma$ error band of the continuum
   values has been indicated by the arrows.}
\label{fg.Cv} \end{figure}

\begin{figure}
\begin{center}
   \scalebox{0.6}{\includegraphics{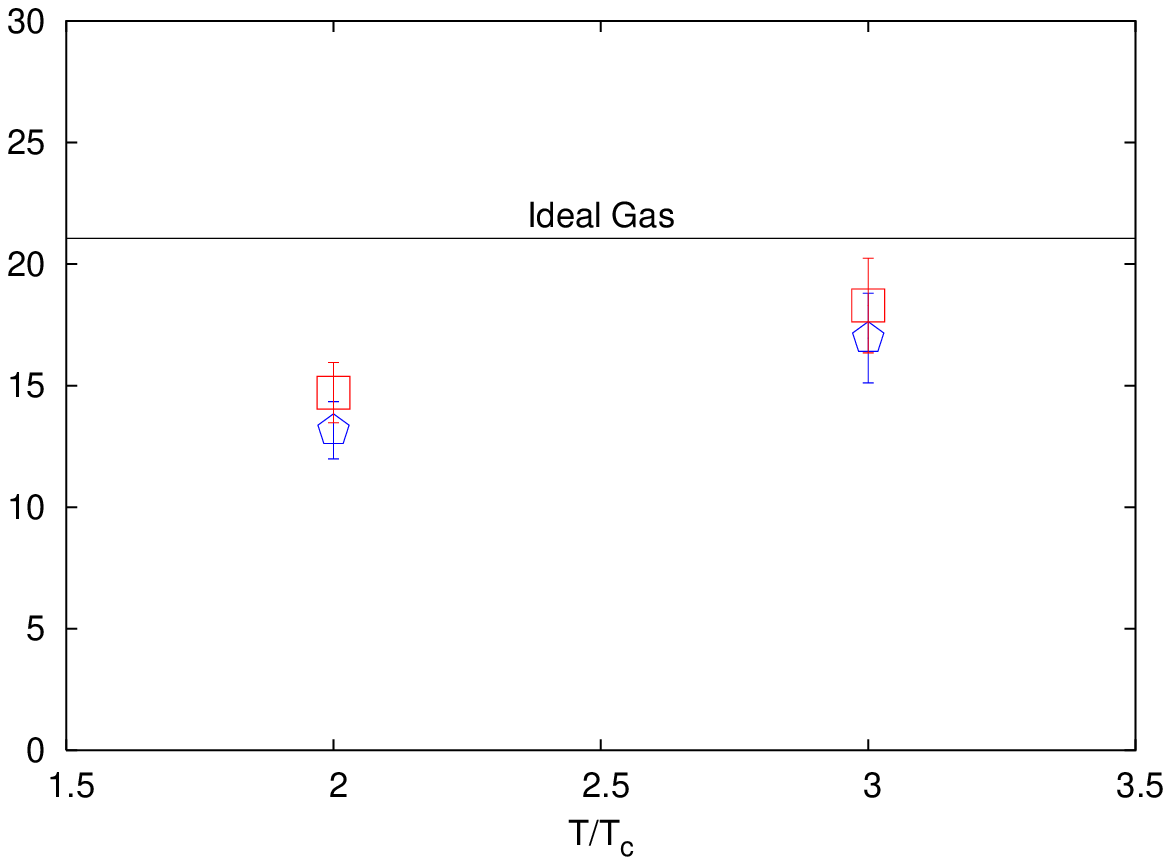}}
   \scalebox{0.6}{\includegraphics{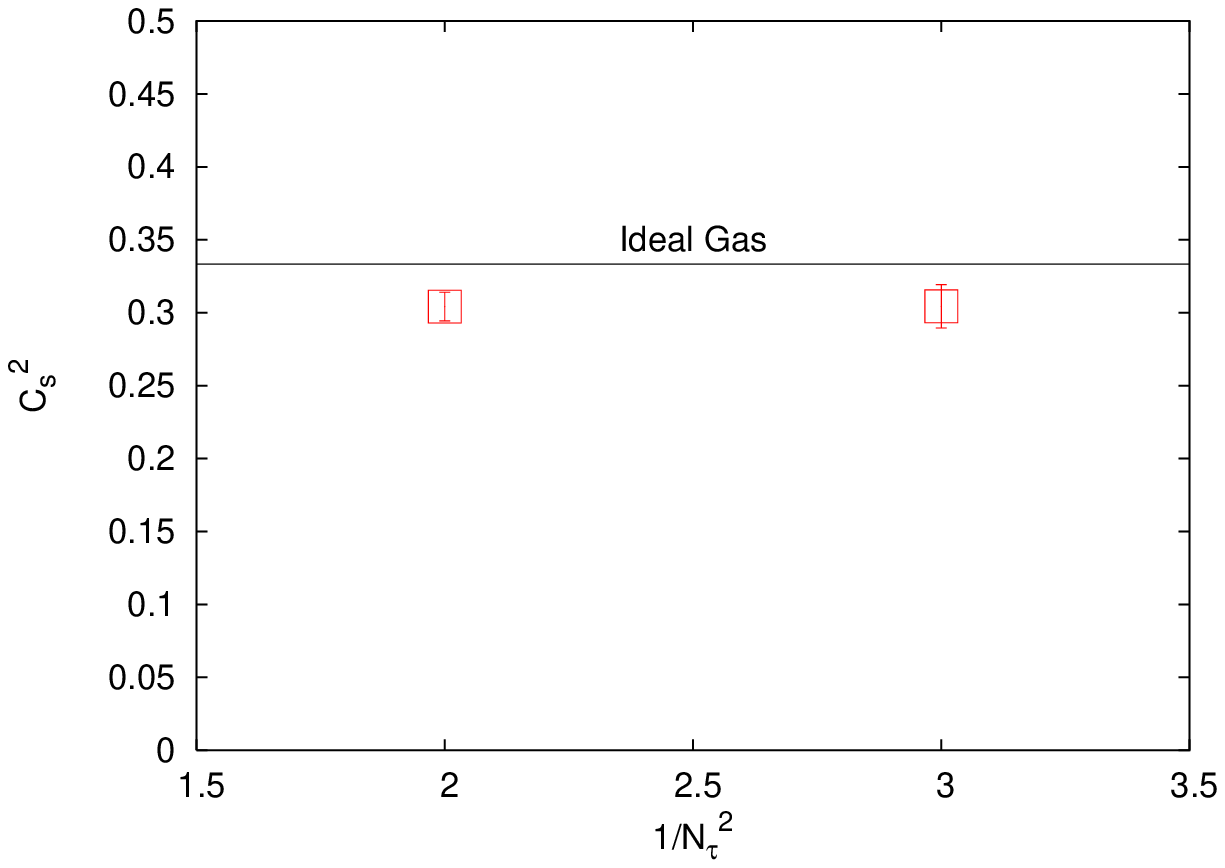}}
\end{center}
\caption{In the left panel we show the continuum values of $4\epsilon/T^4$
   (pentagons) and $\cv/T^3$ (squares) against $T/T_c$. In the right panel
   we show the temperature dependence of the continuum extrapolated values
   of $\cs^2$.}
\label{fg.Cs} \end{figure}

\begin{figure}
\begin{center}
   \scalebox{0.6}{\includegraphics{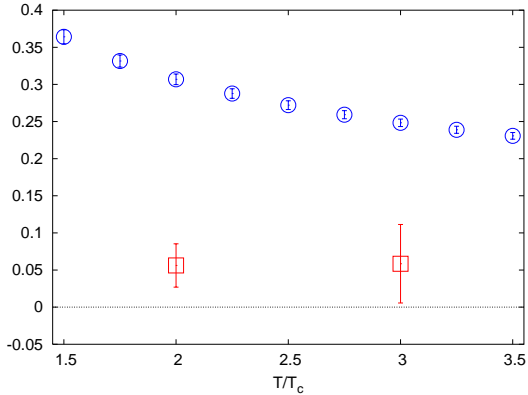}}
\end{center}
\caption{Continuum values of the conformal measure, $\C$, (squares) and the
   one-loop order $\beta$-function (circles) evaluated at the scale $2\pi T$.
   For the $T=0$ gluon condensate see the text.}
\label{fg.D-E} \end{figure}

\section{Simulations and results} \label{sc.results}

The simulations have been performed using the Cabbibo-Marinari
pseudo-heatbath algorithm with Kennedy-Pendleton updating of three $SU(2)$
subgroups on each sweep. Plaquettes were measured on each sweep. For each
simulation we discarded around 5000 initial sweeps for thermalization. We
found that integrated autocorrelation time for the plaquettes never
exceeded $3$ sweeps. In Table \ref{tb.simulation} we give the details
of our runs.  All errors were calculated by the jack-knife method,
where the length of each deleted block was chosen to be at
least six times the maximum integrated autocorrelation time of all the
simulations used for that calculation.

Of all the quantities which go into determining the equation of
state, namely $\Ds$, $\Dt$ and $\Ds-\Dt$, we found that $\Ds$ was the
smallest. Hence control over the errors of $\Ds$ was the most stringent
requirement on the amount of statistics needed. In Figure \ref{fg.Ds-stat}
we show the stability of $\Ds$ against the finite temperature and zero
temperature statistics for the case where we have minimum statistics,
namely for the $3T_c$ run on the $12\times38^3$ lattice and the run at
the same coupling on the $38^4$ lattice. Note that the plaquette values
are of the order of unity, and the first four digits cancel in computing
$\Ds$. Thus, the control over errors shown in Figure \ref{fg.Ds-stat}
was due to our reducing the errors in the plaquette variables to
a few parts per million.

We checked the stability of $\Ds$ against the spatial size
of the $T>0$ lattice. The left panel of Figure \ref{fg.Ds-vol} displays
the dependence of $\Ds$ on $N_s$ for $N_\tau=4$ when the temperature is $2T_c$
and the symmetric lattice size is $22^4$. We have also shown a fit to a 
constant using the data on the three largest lattices. These fits pass
through the data collected on the $N_s=(T/T_c)N_\tau +2$ lattice. We have 
checked that this condition holds for $3T_c$ also. This observation is 
consistent with the results of earlier investigation of
finite size effects for $T>0$ \cite{saumen} and motivated our choice of
$N_s=(T/T_c) N_\tau+2$.

We also investigated the dependence of $\Ds$ on the size of the symmetric 
$N_s^4$ lattice used for the $T=0$ subtraction. The right panel of Figure
\ref{fg.Ds-vol} exhibits the dependence of $\Ds$ on $N_s$ for a run at
$2T_c$ on a $8\times 18^3$ lattice. $\Ds$ is seen to be constant for
$N_s\ge22$. In view of this we have used $22^4$ as our minimum $T=0$
lattice size, and scaled this up with changes in the lattice spacing.

We evaluated the contributions of the terms containing different 
covariances of the plaquettes and found them to be
totally negligible, as shown in Table \ref{tb.covar}.
It is worth noting that a previous computation
in $SU(2)$ pure gauge theory close to $T_c$ found significantly larger,
and clearly non-vanishing, variances of the plaquettes \cite{gavai}. This
suggests that the variance terms might make significant contributions
to the specific heat and speed of sound closer to the softest point of
the EOS.

Following the observation illustrated in Figure \ref{fg.D-int-diff},
all our continuum extrapolations have been done with lattice spacings
which are at least as small as that at $\beta=6.55$. For $T=2T_c$ this
leaves three values of $N_\tau$ from which an continuum extrapolation
linear in $a^2\propto1/N_\tau^2$ can be performed.  However, at $T=3
T_c$, the continuum extrapolation has been performed with two values
of $N_\tau$. This completely fixes the two parameters of the linear
extrapolation to the continuum and the error in the continuum extrapolated
value is by definition zero. In this case the error in the continuum
value was estimated by three methods. Two of these consisted of first
making the best fit using two parameters, and then keeping one fixed while
allowing the other to vary in order to make an estimate of the error in
that parameter.  We also made extrapolations using the upper end of one
error bar and the lower end of the other. This last procedure gave the
maximum errors in the continuum extrapolated values, and we choose to
quote this, since it is the most conservative error estimate.

In Figure \ref{fg.E-P} we present our results for $\epsilon$ and $P$. At
$3T_c$ the continuum limit values of $\epsilon /T^4$ and $P/T^4$ differ
from their respective ideal gas values ($8\pi^2/15$ and $8\pi^2/45$
respectively) by about 24\%, and by about 40\% at $2T_c$. This is
consistent with previous measurements of the equation of state at
these temperatures.  At $2T_c$ our results differ from the ideal gas
value by almost 7-$\sigma$. The actual results are also collected in
Table \ref{tb.cont-values}.

Similar continuum extrapolation of $\cv/T^3$ and $\cs^2$ are shown
in Figures \ref{fg.Cv} and Figures \ref{fg.Cs} respectively. Continuum
extrapolated values of different quantities for two temperatures are
tabulated in Table \ref{tb.cont-values}. It can be seen that although
$\epsilon /T^4$ and $P/T^4$ differ significantly from their ideal gas
values, $\cs^2$ is quite close to $1/3$, being within 2-3 $\sigma$. 
Similarly, it can be seen that $\cv/T^3$ is completely compatible with 
$4\epsilon/T^4$, a fact also illustrated in Figure \ref{fg.D-E}.  However, 
its deviation from the ideal gas value is seen to be more significant.

The generation of a scale and the consequent breaking of conformal
invariance at short distances in QCD is quantified by the $\beta$-function
of QCD, and at long distance in the finite temperature plasma by the
conformal measure $\C$.  In Figure \ref{fg.D-E} we plot these two
together. The $T=0$ gluon condensate, $\langle F^2\rangle_0$ gives a
measure of the non-perturbative strength of the breaking of conformal
invariance at $T=0$. A recent estimate \cite{badalian} gives $\langle
F^2\rangle_0 / \lambdams^4=10.5\pm0.5$, where $\lambdams$ is compatible
with the value used to set the temperature scale.  It is clear that
conformal invariance is better respected in the finite temperature
effective long-distance theory than at the microscopic scale.

An alternative measure of conformal symmetry breaking for $T>0$ could
be the dimensionless quantity
\beq
   \C' \equiv \frac{\Delta}{\langle F^2\rangle_0} =
      \C \frac{\epsilon}{\langle F^2\rangle_0} =
      \C\left(\frac{\epsilon}{T^4}\right)
      \left(\frac T{\lambdams}\right)^4
      \left(\frac{\langle F^2\rangle_0}{\lambdams^4}\right)^{-1}.
\label{altc}\eeq
Clearly its use is limited to making the trivial point that the
scale $T$ breaks conformal symmetry explicitly, reflected by the $T^4$
growth of $\C'$.  This misses physics on both the weak-coupling and the
strong coupling AdS/CFT sides. On one side is the likelihood that the
weak coupling expansion, which is conformal, becomes more accurate at
sufficiently high temperature--- although $\Delta$ increases with $T$,
$\Delta/T^4$ becomes a small part of $\epsilon/T^4$, $P/T^4$ or $s/T^4$.
On the other hand is the possibility that AdS/CFT is accurate because
the $T$ dependence can be taken into the dimension of operators, and
the CFT shows up in non-trivial dimensionless quantities such as the
function $f$ in eq.\ (\ref{sym}).  $\C$ allows us to test the accuracy
of these possibilities, thereby capturing the interesting facts shown in
Figure \ref{fg.Cs}, whereas $\C'$ misses them.

\begin{table}
\begin{center}
\begin{tabular}{|c|c||c|c|c||c|c|} \hline
$T/T_c$&$g^2 N_c$&$\epsilon/T^4$&$P/T^4$&$s/T^3$&$\cv/T^3$&$\cs^2$\\ \hline
2.0&8.1 (1)&3.3 (3)&1.0 (1)&4.2 (3)&15 (1)&0.30 (1)\\ \hline
3.0&7.0 (1)&4.2 (5)&1.3 (2)&5.6 (6)&18 (2)&0.30 (2) \\ \hline
\end{tabular}
\end{center}
\caption{Continuum values of some quantities at the two temperatures
   we have explored. The numbers in brackets are the error on the
   least significant digit. The value of the t'Hooft coupling $N_c g^2$
   is computed at the scale $2\pi T$ using the $T_c/\lambdams$ quoted
   in \cite{sgupta}.}
\label{tb.cont-values} \end{table}

\begin{figure}
\begin{center}
   \includegraphics{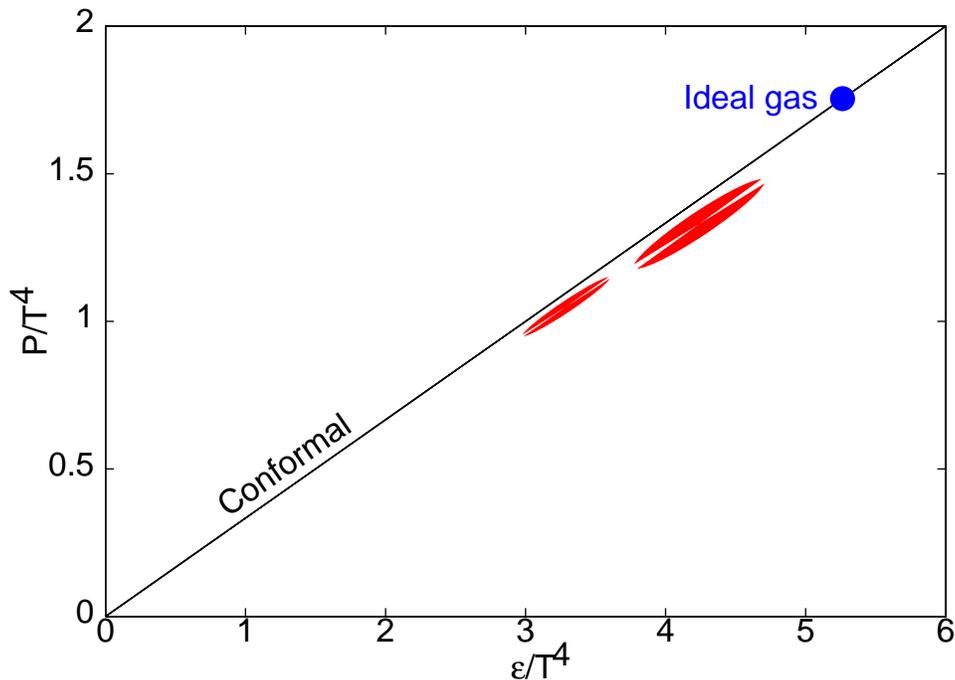}
\end{center}
\caption{The equation of state of QCD matter. The diagonal line denotes possible
    EOS for theories with conformal symmetry. The circle on the diagonal denotes
    the ideal gluon gas, whose EOS in this form is temperature independent. The
    ellipses denote 66\% error bounds on the measured EOS. The ratio of the
    axes is a measure of the covariance in the measurements of $\epsilon/T^4$
    and $P/T^4$, which is about 90\%. The wedges piercing these ellipses have
    average slope $\cs^2$, and the opening half-angle of these wedges indicate
    the error in $\cs^2$.}
\label{fg.eos} \end{figure}

\section{Discussion} \label{sc.summary}

In this paper we have determined the continuum limits of the specific heat
at constant volume, $\cv$, and the speed of sound, $\cs$, in the continuum
limit of the pure gluon plasma. In the process we have also recomputed
the EOS, \ie, the pressure, $P$, and the energy density, $\epsilon$, by a
method which has not been used earlier to obtain the continuum limit. Our
results are collected together in Table \ref{tb.cont-values}. It is clear
from this that the plasma phase is not far from the conformal symmetric
limit, in which $\cs^2=1/3$ and $\cv/T^3=4\epsilon/T^4$. This is also
illustrated in Figures \ref{fg.Cs} and \ref{fg.D-E}. The latter figure
further shows that conformal symmetry is more nearly realized on length
scale of order $T$ than one would expect either from the $\beta$-function of QCD
or from the gluon condensate $\langle F^2\rangle_0$.

We have extended the operator method for the EOS in order to obtain these
results. We have limited our computations to lattice spacings where the
method is guaranteed to work because of the observed scaling of results
with the correct QCD beta function, as shown in Figure \ref{fg.D-int-diff}.
The main advantage of this method is the relatively noise-free determination
of $\cs$ and $\cv$.

There is as yet no unambiguous determination of these quantities from
experiments. Hydrodynamic analyses of RHIC data routinely use
$\cs^2=1/3$ in the plasma phase; our results show that this receives a
correction of up to 10\% in QCD. It would be interesting to check the
sensitivity of the RHIC data to the variation in $\cs^2$ and compare the
bounds there with our lattice results.  Little is known yet about $\cv$.
The only estimate that we are aware of \cite{korus} uses the data on
$\pt$ fluctuations from \cite{appelshauser} to obtain $\cv/T^3=60\pm100$
for $T=180$ MeV.

Our data on the entropy density, $s/T^3$, shown in Table
\ref{tb.cont-values} can be used to test the quantitative predictions
of the strong-coupling expansion of the ${\cal N}=4$ supersymmetric Yang-Mills
theory in \cite{gubser}.  Eq,\ (\ref{sym}) should interpolate the ratio $s/s_0$
between the strong-coupling limit of 3/4 and the weak-coupling limit
of 1. It is clear that at $T=2T_c$ this ratio is below 3/4, and hence
cannot be described by the computation in \cite{gubser}. At $3T_c$,
the ratio is above 3/4 and there is quantitative agreement with the
formula of \cite{gubser} using the value of the t'Hooft coupling quoted
in Table \ref{tb.cont-values}.  It is interesting that this particular
strong-coupling theory fails at temperatures where weak coupling theories
also break down.

A partial summary of our results is illustrated by plotting the equation
of state as $P/T^4$ against $\epsilon/T^4$, as in Figure \ref{fg.eos}. In
this plot, the ideal gas for fixed number of colours is represented by
a single point which is independent of $T$, and theories with conformal
symmetry by the line $\epsilon=3P$. Pure gauge QCD lies close to the
conformal line at high temperature, as shown. One expects the EOS to drop
well below this line near $T_c$, since the theory then contains massive
hadrons (glueballs in pure gluon QCD) with masses in excess of $T_c$. Thus
one expects the origin to be approached almost horizontally as $T\to0$. We
will present data closer to, and below, $T_c$ in a forthcoming paper.

SG would like to thank Jean-Paul Blaizot, Kari Eskola, Rajesh Gopakumar,
Pasi Huovinen, Toni Rebhan and Vesa Ruuskanen for discussions. RVG
would like to thank the Alexander von Humboldt Foundation for financial
support and the members of the Theoretical Physics Department of Bielefeld
university for their kind hospitality. SM would like to thank Rajarshi
Ray for discussions and for his help in calculating the derivatives of
Karsch Coefficients.  SM would also like to thank TIFR Alumni Association
for partial financial support.

\appendix

\section{Derivatives of F and G} \label{sc.variance-terms}

From the definition of $F$ in eq.\ (\ref{F-G}) we obtain
\beq
   \left.\frac{\partial F}{\partial \xi}\right|_{a_s} =
      a_s \left[ \frac{\partial K_s'}{\partial a_s}\Ds
             + \frac{\partial K_{\tau}'}{\partial a_s}\Dt \right]
    + a_s \left[ \frac{\partial K_s}{\partial a_s}\Ds'
             + \frac{\partial K_{\tau}}{\partial a_s}\Dt' \right].
\eeq
Since the plaquette operators do not explicitly depend on $\xi$ (or $a_s$), we
can easily take the derivative---
\beq
   D_i'=D_i\langle S'\rangle-\langle(P_i-P_0)S'\rangle  
     = -6 N_c N_\tau N_s^3 \left[ K_s'\sigma_{s,i}+K_\tau'\sigma_{\tau,i} \right],
\eeq
where $\sigma_{s,i}=\langle\Ds D_i\rangle-\langle\Ds\rangle\langle D_i\rangle$ and
$\sigma_{\tau,i}=\langle\Dt D_i\rangle-\langle\Dt\rangle\langle D_i\rangle$ are the
variances and covariances of the plaquettes. Next, using Eq~(\ref{coup-der-1}) and
Eq~(\ref{coup-der-2}), the $\xi \rightarrow 1$ limit becomes
\beq
   \left.\frac{\partial F}{\partial\xi}\right|_{a_s} = \frac{B(\alpha_s)}{2 \pi \alpha_s^2} \left[\Dt-\Ds\right]
     -6N_cN_\tau N_s^3 \frac{B(\alpha_s)}{2 \pi \alpha_s^2} \left[ \frac{\sigma_{\tau,\tau}-\sigma_{s,s}}{g^2}
           +c_s'\sigma_{s,s}+c_{\tau}'\sigma_{\tau,\tau}
              +(c_s'+c_\tau')\sigma_{s,\tau}\right].
\eeq
Similarly we find
\beqa
\nonumber
   \left.\frac{\partial G}{\partial \xi}\right|_{a_s} &=&
     \frac{\Ds+\Dt}{g^2}-c_s^{\prime} \Ds + 3 c_{\tau}^{\prime} \Dt 
      + c_s^{\prime \prime} \Ds  + c_{\tau}^{\prime \prime} \Dt
    -6N_cN_\tau N_s^3 \times\\
   &&\qquad\left[ 
       \frac{\sigma_{s,s}+\sigma_{\tau,\tau}-2\sigma_{s,\tau}}{g^4}
       +\frac{2(c_{\tau}'\sigma_{\tau,\tau}+c_s'\sigma_{s,\tau}-c_s'\sigma_{s,s}
             -c_{\tau}'\sigma_{s,\tau})}{g^2}
       +{c_s'}^2\sigma_{s,s}+{c_\tau'}^2\sigma_{\tau,\tau}+2c_s'c_\tau'\sigma_{s,\tau}
         \right].
\eeqa
Since $D_i\sim g^2$ and $\sigma_{i,j}\sim g^4$ in the limit $g\to0$,
$\partial F/\partial\xi\to0$ but $\partial G/\partial \xi$ remains
finite. As a result, $\Gamma\to0$, in this limit, as it should, even
when the variances of the plaquettes are taken into account.

\section{Second order Karsch coefficients} \label{sc.karsch-coeff} 

Expressions for $c_s(\xi)$ and $c_{\tau}(\xi)$ as given in \cite{Kar} are
\beqa
\nonumber
   c_s(\xi) &=& 4N_c\left[ \frac{N_c^2-1}{24N_c^2} \left(I_1(\xi)-\frac34\right)
        -\frac5{288}I_{2a}(\xi) + \frac1{48}I_3(\xi) + \frac1{128}I_4(\xi)
        +\frac{11}{12}FIN(\xi) + 0.010245 \right ], \\
\nonumber
   c_{\tau}(\xi) &=& 4N_c\left[ \frac{N_c^2-1}{24N_c^2} \left(\frac1{3\xi^2}I_1(\xi)
        +\frac1{\xi}I_5(\xi)-\frac12\right) + \frac1{64}I_3(\xi) 
        - \frac5{576}I_{2a}(\xi) \right] \\ 
   &&\qquad
        + 4N_c\left[ \frac1{256\xi^2}I_4(\xi) - \frac1{48\xi^2}I_6(\xi)
        - \frac1{192\xi^2}I_7(\xi) + \frac{11}{12}FIN(\xi) + 0.010245 \right]. 
\eeqa
Here $b^2=\sin^2x_1+\sin^2x_2+\sin^2x_3$, and
\beqa
\nonumber
   I_1(\xi) &=& \xi\left(\frac2\pi\right)^3\int d^3xb(\xi^2+b^2)^{-1/2},\\
\nonumber
   I_{2a}(\xi) &=& \xi\left(\frac2\pi\right)^3\int
                               d^3xb^{-1}(\xi^2+2b^2)(\xi^2+b^2)^{-3/2},\\
\nonumber
   I_{2b}(\xi) &=& \xi^3\left(\frac2\pi\right)^3\int
      d^3x\left[b(\xi^2+b^2)^{1/2}\left(b+(\xi^2+b^2)^{1/2}\right)^2\right]^{-1},\\
\nonumber
   I_3(\xi) &=& \xi\left(\frac2\pi\right)^3\int d^3x\sin^2x_1\sin^2x_2
      \frac{\xi^2+2b^2}{b^3(\xi^2+b^2)^{3/2}},\\
\nonumber
   I_4(\xi) &=& \xi\left(\frac2\pi\right)^3\int d^3x\sin^22x_1
      \frac{\xi^2+2b^2}{b^3(\xi^2+b^2)^{3/2}},\\
\nonumber
   I_5(\xi) &=& \xi^2\left(\frac2\pi\right)^3\int d^3x(\xi^2+b^2)^{-1/2}
      \left[b+(\xi^2+b^2)^{1/2}\right]^{-1},\\
\nonumber
   I_6(\xi) &=& \xi^3\left(\frac2\pi\right)^3\int
      d^3xb^{-1}(\xi^2+b^2)^{-3/2}\cos^2x_1,\\
   I_7(\xi) &=& \xi^3\left(\frac2\pi\right)^3\int d^3xb^{-1}(\xi^2+b^2)^{-3/2},\\
\eeqa
Limits of all the above integrals are $[ 0, \pi/2 ]$. And :
\beqa
\nonumber
   DIV(\xi) &=& \frac1{(2\pi)^4}\int\int_{-\pi/2}^{\pi/2}\int d^3x
     \int_{-\pi\xi/2}^{\pi\xi/2} dx_4\left[b^2+\xi^2\sin^2(x_4/\xi)\right]^{-2},\\
   FIN(\xi) &=& DIV(\xi)-DIV(1).
\eeqa

The integral $DIV(\xi)$ is infrared divergent. However, what actually needed 
here is $FIN(\xi)$. $FIN(\xi)$ is not divergent as it is constructed by 
subtracting the integral $DIV(1)$, having the same infrared divergence as 
that of $DIV(\xi)$, from $DIV(\xi)$. The derivatives of $FIN(\xi)$ 
are just the derivatives of $DIV(\xi)$ which are not divergent. We calculated
 $FIN^{\prime\prime}(1)$ by direct numerical integration and also by taking 
a derivative of $FIN^{\prime}(\xi)$, at $\xi=1$, numerically. We found that 
values obtained from both the methods are consistent. 

\begin{table}
\begin{center}
\begin{tabular}{|c|c c c c c c c c c|}
\hline 
&$I_1$&$I_{2a}$&$I_{2b}$&$I_3$&$I_4$&$I_5$&$I_6$&$I_7$&$FIN$ \\
\hline
Integrals&0.750000&0.929600&0.119734&0.103289&0.478934&0.250000&0.206578&0.309867&0.0\\
\hline
$1$-st derivatives&0.440133&0.208546&0.190133&0.033774&0.065779&0.309867&0.238384&0.411188&0.003166 \\
\hline
$2$-nd derivatives&-0.518412&-0.663270&0.030921&-0.088777&-0.262319&-0.101321&-0.146561&-0.159106&-0.014471 \\
\hline
\end{tabular}
\end{center}
\caption{Values of the integrals $I_x(\xi)$ 's  and $FIN(\xi)$ and their derivatives with respect to $\xi$ at $\xi = 1$.}
\label{tb.I-values} \end{table}

The numerical values of all these above integrals and their derivatives
with respected to $\xi$, at $\xi=1$, are tabulated in Table
\ref{tb.I-values}. Using these values we can obtain $\xi$ derivatives
of the Karsch coefficients at $\xi=1$
\beqa
\nonumber
   c_s' = 4N_c\left[ \frac{N_c^2-1}{32N_c^2}0.586844+0.000499\right],\\
\nonumber
   c_s'' = 4N_c\left[ \frac{1-N_c^2}{32N_c^2}0.691216-0.005649\right],\\
\nonumber
   c_\tau' = 4N_c\left[ \frac{1-N_c^2}{32N_c^2}0.586844+0.005306\right],\\
   c_\tau'' = 4N_c\left[ \frac{N_c^2-1}{32N_c^2}1.038595-0.001044\right].
\eeqa
The values of all the integrals and their first derivatives as well
as the regular Karsch coefficients match with their respective values
mentioned in \cite{Kar}.

\end{document}